\documentclass{article}
\usepackage{amsmath,amssymb,amscd,latexsym,amsthm}
\usepackage{graphicx}
\usepackage{slashed}
\usepackage{hyperref}
\usepackage[backend=bibtex]{biblatex}
\usepackage{tikz}
\usepackage{tikz-cd}
\author{Peter Woit \\
Department of Mathematics, Columbia University\\
woit@math.columbia.edu}
\title{Notes on the Twistor $\mathbf P^1$} 
\begin{document}
\maketitle
\begin{abstract}
Remarkably, the twistor $\mathbf P^1$ occurs as a fundamental object in both four-dimensional space-time geometry and in number theory.   In Euclidean signature twistor theory  it is how one describes space-time points.  In recent work by Fargues and Scholze on the local Langlands conjecture using geometric Langlands on the Fargues-Fontaine curve, the twistor $\mathbf P^1$ appears as the analog of this curve at the infinite prime.

These notes are purely expository, written with the goal of explaining, in a form accessible to both mathematicians and physicists, various different ways in which the twistor $\mathbf P^1$ makes an appearance, often as a geometric avatar of the quaternions. 
\end{abstract}

\section{Introduction}

In the Euclidean signature twistor description of space-time, a point in space-time is described by a \lq\lq twistor $\mathbf P^1$":  the Riemann sphere, together with the real structure given by the antipodal involution.  The physical interpretation of this sphere is that it corresponds to the sphere of light-rays through the corresponding point.   Remarkably, the same twistor $\mathbf P^1$ occurs not just in this physical context, but in two related mathematical contexts far removed from physics:
\begin{itemize}
\item In Simpson's approach to Hodge theory, Hodge structures are given by equivariant vector bundles on the twistor $\mathbf P^1$.
\item Recent work in arithmetic geometry and the Langlands program has made clear that a useful tool for understanding behavior at a prime $p$ is a new sort of geometrical object, the Fargues-Fontaine curve.   The twistor $\mathbf P^1$ is the analog of the Fargues-Fontaine curve at the \lq\lq infinite prime" (the archimedian place of the number field $\mathbf Q$).
\end{itemize}

Euclidean twistor geometry provides a compelling basis for unification of the fundamental internal and space-time symmetries of physics (see \cite{woit}).  That the same twistor $\mathbf P^1$ describing  points of physical space-time there also fits into a description of the points of number theory  is a remarkable relation between these two very different subjects.   The goal of these notes is to gather together some more detail  about this observation, the significance of which remains unclear.  Along the way, we'll also point to other occurrences of the same twistor $\mathbf P^1$ in such different contexts as hyperk\"ahler geometry and the metaplectic central extension of the symplectic symmetry group of canonical quantization.

\section{The twistor $\mathbf {P}^1$}

The two-dimensional  Riemann sphere can be identified with $\mathbf{CP}^1$, the space of complex lines in $\mathbf C^2$.    Homogeneous coordinates 
$$[z_1,z_2]$$
 describe points of this space as equivalence classes of pairs of complex numbers.   Taking the quotient
$$z=z_1/z_2$$ gives a coordinate chart identifying $\mathbf C$ with $\mathbf{CP}^1$ minus the point $[z_1,0]$.

A real structure on $\mathbf{CP}^1$ is given by an antiholomorphic map
$$\rho: \mathbf{CP}^1\rightarrow \mathbf{CP}^1$$
such that 
$$\rho^2=1$$
One such real structure is the usual complex conjugation.  In homogeneous coordinates
$$\rho([z_1,z_2])=[\overline{z}_1,\overline{z}_2]$$
or in the $z$ coordinate
$$\rho (z)=\overline{z}$$
Real points for this real structure on $\mathbf{CP}^1$ are those satisfying 
$$\rho([z_1,z_2])=[z_1,z_2]$$
so are on a circle in $\mathbf{CP}^1$, or the real number line in the $z$ coordinate.

The twistor $\mathbf{P}^1$ (which we'll often denote $\mathbf P_{tw}^1$) is $\mathbf{CP}^1$ with a different real structure, given in homogeneous coordinates by
$$\rho_{tw}([z_1,z_2])=[-\overline{z}_2,\overline{z}_1]$$
or in the $z$ coordinate by
$$\rho_{tw}(z)=-1/{\overline z}$$
As a map on coordinates $(z_1,z_2)$ of $\mathbf C^2$ one has $\rho_{tw}^2=-1$, but $\rho_{tw}^2=1$ as a map on the space of complex lines.
On the sphere, $\rho_{tw}$ is the antipodal map taking a point to its antipode.   Note that there are no points on $\mathbf{CP}^1$ such that 
$$\rho_{tw}([z_1,z_2])=[z_1,z_2]$$
since 
$$z=-1/{\overline z}\implies |z|^2=-1$$

As an object in algebraic geometry, one way the twistor $\mathbf P^1$ can be described is as above, as a complex projective variety ($\mathbf{CP}^1$) together with an action of the Galois group $Gal(\mathbf C/\mathbf R)$.  It can also be defined in terms of the equation $x^2+y^2+z^2=0$, which describes a curve in projective two-space.  This is a projective  algebraic curve defined over the real numbers, with complex points but no real points.

One way to understand where $\rho_{tw}$ comes from is in terms of the quaternions $\mathbf H$.  A conventional identification of $\mathbf H$ with $\mathbf C^2$  is given by
$$x_0+x_1\mathbf i +x_2\mathbf j + x_3\mathbf k \in \mathbf H \leftrightarrow (x_0+x_1i, x_2+x_3i)\in \mathbf C^2$$ 
allowing one to write quaternions as (identifying $i\in \mathbf C^2$ and $\mathbf i\in \mathbf H$)
$$q=z_1+ z_2\mathbf j$$ 
for complex numbers $z_1=x_0+x_1i$ and $z_2=x_2+x_3i$. $\rho_{tw}$ is then multiplication by $\mathbf j$, since  
\begin{align*}
\mathbf j (z_1+ z_2\mathbf j)=&\mathbf j(x_0+x_1\mathbf i) +\mathbf j(x_2 + x_3\mathbf i)\mathbf j\\
=&-(x_2-x_3\mathbf i)+(x_0-x_1\mathbf i)\mathbf j\\
=&-\overline z_2 +\overline z_1 \mathbf j
\end{align*}

\section{Twistor theory in (compactified) Euclidean space-time}

Penrose's twistor theory generalizes the use of $\mathbf C^2$ and $\mathbf H$ above to the case of $\mathbf C^4$ and $\mathbf H^2$.   $T=\mathbf C^4$ is Penrose's twistor space, and instead of $\mathbf {CP}^1$ one considers the projective twistor space $\mathbf {CP}^3=PT$ of complex lines in $T$.   In twistor theory, a point in (complexified, compactified) space-time is a $\mathbf C^2\subset T$, so a point in the Grassmannian $Gr_{2,4}(\mathbf C)$.  This provides a tautological definition of the (Weyl) spinor bundle: the Weyl spinors at a point are the point.  In terms of $PT$, a space-time point is a $\mathbf {CP}^1\subset PT$.  This $\mathbf {CP}^1$ is the sphere of light rays through the corresponding point, exactly the sphere one sees when one opens an eye.

Most discussions of twistor theory and physics focus on a real four-dimensional Minkowski signature subspace of $Gr_{2,4}(\mathbf C)$, but one can relate physics there by analytic continuation to what happens on a real four-dimensional Euclidean signature subspace. To define such a subspace, one identifies $\mathbf C^4$ and $\mathbf H^2$ using a pair of the earlier identifications of $\mathbf C^2$ with $\mathbf H$ by $z_1+z_2\mathbf j$.  As in the $\mathbf {CP}^1$ case, one has two different real structures on $\mathbf {CP}^3$, given by the usual conjugation
$$\rho([z_1,z_2,z_3,z_4])=[\overline{z}_1,\overline{z}_2,\overline{z}_3,\overline{z}_4]$$
and by multiplication by $\mathbf j$
$$\rho_{tw}([z_1,z_2,z_3,z_4])=[-\overline{z}_2,\overline{z}_1,-\overline{z}_4,\overline{z}_3]$$

$\mathbf {CP}^3$ with the real structure $\rho$ has a three-real-dimensional space $\mathbf {RP}^3$ of real points, but the twistor real structure $\rho_{tw}$ has no real points. It does however have real lines, i.e. $\mathbf {CP}^1$s in $\mathbf {CP}^3$ that are fixed by $\rho_{tw}$.  Each such $\mathbf {CP}^1$ is a twistor $\mathbf P^1$, with $\rho_{tw}$ acting on it in the manner described in the previous section.   There will be a four-real-dimensional family of such $\mathbf {CP}^1$s, parametrized by the four-sphere $S^4$.  To see this, consider the quaternionic analog $\mathbf {HP}^1$ of $\mathbf {CP^1}$, with homogeneous coordinates
$$[q_1,q_2]$$
which describe points on this space as equivalence classes of pairs of quaternions.   Taking the quotient
$$q=(q_2)^{-1}q_1$$ gives a coordinate chart identifying $\mathbf H$ with $\mathbf{HP}^1=S^4$ minus the point $[q_1,0]$.

Our identification of $\mathbf C^4$ and $\mathbf H^2$ gives a mapping
$$\pi: [z_1,z_2,z_3,z_4]\in \mathbf{CP}^3\rightarrow [z_1+z_2\mathbf j,z_3+z_4\mathbf j]\in \mathbf {HP}^1$$ 
that takes a complex line in $\mathbf C^4$ to the quaternionic line in $\mathbf H^2$ that it generates.  This mapping fibers $\mathbf {CP}^3$ over $S^4$, with fibers the $\mathbf {CP}^1$s that are real lines for the real structure $\rho_{tw}$.

For the case of $\mathbf C^2$, from the linear action of $SL(2,\mathbf C)$ on $\mathbf C^2$ one gets an action of $SL(2,\mathbf C)$ on $\mathbf{CP}^1$ by conformal transformations, an action which does not commute with the real structure.  In the case of $\mathbf C^4=\mathbf H^2$, one gets an action of the conformal group $SL(2,\mathbf H)=Spin(5,1)$, again not commuting with the real structure.

There are only the two inequivalent real structures $\rho$ and $\rho_{tw}$ on projective twistor space $\mathbf{CP}^3$.    To pick out (compactified) Minkowski space-time points in the complexification $Gr_{2,4}(\mathbf C)$ one needs to use a real structure that does not act on $PT=\mathbf{CP}^3$ but instead takes $PT$ to a dual projective space $PT^*$.  In terms of spinors, this Minkowski space real structure takes one chirality of spinors to the other (unlike the Euclidean case).

\section{Twistor theory for $\mathbf R^4$}

The $\mathbf {CP}^1$ fiber in the fibration
\begin{equation*}
	\label{eq:twistorfibration}
	\begin{tikzcd}
		\mathbf {CP}^1 \arrow[r]& PT=\mathbf {CP}^3 \arrow[d,"\pi"]\\
		& S^4=\mathbf {HP}^1
	\end{tikzcd}
\end{equation*}
of the previous section can be interpreted as parametrizing orthogonal complex structures on the tangent space of the base space, allowing one to study $S^4$ (which is not a complex manifold) using $\mathbf{CP}^3$, which is.  Solutions of the self-dual Yang-Mills equations on $S^4$ correspond to holomorphic structures on vector bundles on $\mathbf {CP}^3$, and this can be generalized from $S^4$ with its usual metric to self-dual solutions to the Einstein equations on other four-manifolds (this is Penrose's non-linear graviton construction\cite{nonlinear-graviton}).

One can instead pursue the idea of working with all orthogonal complex structures on $\mathbf R^4=\mathbf H$ simultaneously by identifying such complex structures with points on the sphere of unit length imaginary quaternions
$$x_1\mathbf i +x_2\mathbf j +x_3\mathbf k$$
where $\mathbf x=(x_1,x_2,x_3)$ is a unit vector.
This sphere of complex structures on $\mathbf R^4=\mathbf H$ can  be identified with $\mathbf {CP}^1$ and given the usual complex structure on that space.

Instead of Penrose's twistor space $T$ and its projective version $PT$,  one can form the product of $\mathbf R^4$ and $\mathbf{CP}^1$, with the fibration
\begin{equation*}
	\begin{tikzcd}
		 &\mathbf R^4\times \mathbf {CP}^1 \arrow[d,"\pi"]\\
		& \mathbf {CP}^1
	\end{tikzcd}
\end{equation*}
where $\pi$ is projection on the second factor.  In this context, $\mathbf R^4\times \mathbf {CP}^1$ is called the twistor space.  It can be given the complex structure that on $\mathbf R^4$ is determined by the point on $\mathbf {CP}^1$ and on $\mathbf{CP}^1$ is the usual complex structure.   With respect to this complex structure, $\pi$ is a holomorphic map.  Note that this construction comes with a real structure $\rho$ which on $\mathbf R^4=\mathbf H$ is the identity, and is the twistor real structure $\rho_{tw}$ on $\mathbf{CP}^1$.

The holomorphic fibration above is isomorphic to the fibration
\begin{equation*}
	\begin{tikzcd}
		 &\mathcal O(1)\oplus\mathcal O(1)\arrow[d,"\pi"]\\
		& \mathbf {CP}^1
	\end{tikzcd}
\end{equation*}
Instead of using quaternions, one can realize the sphere $\mathbf {CP}^1$ in terms of Pauli matrices, as the set of matrices
$$x_1\sigma_1 + x_2\sigma_2 +x_3\sigma_3$$
with $\mathbf x$ a unit vector.  Taking as fiber above such a point the $+1$ eigenspace of the matrix acting on $\mathbf C^2$ gives the vector bundle $\mathcal O(1)$.  Two copies of this construction give $\mathcal O(1)\oplus \mathcal O(1)$, a vector bundle  with total space $\mathbf R^4\times \mathbf {CP}^1$ and fiber $\mathbf R^4$ identified with $\mathbf C\oplus \mathbf C=\mathbf C^2$.
More generally, if one replaces $\mathbf R^4=\mathbf H$ by any quaternionic vector space of dimension $d$, one gets a holomorphic bundle over $\mathbf{CP}^1$, direct sum of $2d$ copies of $\mathcal O(1)$.

\section{The twistor $\mathbf P^1$ and geometry}

In this section we'll see that the twistor $\mathbf P^1$ appears in geometry in several different guises.

\subsection{Hyperk\"ahler manifolds}

One can generalize the notion of a K\"ahler manifold by considering manifolds $M$ of dimension $4k$, which come not with a single K\"ahler structure $I$, but a sphere of such structures, given by
$$aI+bJ+cK$$
where $J,K$ are also K\"ahler structures, $IJ=K$ and $(a,b,c)$ lie on the unit sphere in $\mathbf R^3$.  Besides the simple example of $\mathbf R^{4k}=\mathbf H^k$, other examples of such manifolds include co-adjoint orbits of complex Lie groups, as well as spaces of holomorphic flat $GL(n,\mathbf C)$ connections or Higgs bundles on a Riemann surface $\Sigma$.  For such  manifolds one can define a corresponding twistor space, in the same manner as was done above for the case of $\mathbf R^{4}$.   These twistor spaces were first studied in \cite{hitchin-rocek}.  For a survey of results about hyperk\"ahler manifolds, see \cite{hitchin-hyperkahler}.

The authors of \cite{hitchin-rocek} showed that one could generalize the notion of symplectic quotient by a group $G$ from the case of K\"ahler manifolds to the hyperk\"ahler case, allowing one to construct new hyperk\"ahler manifolds by taking hyperk\"ahler quotients.  A fundamental source of examples is the space of self-dual  Yang-Mills connections on a bundle on $\mathbf R^4$, where quotienting by the group of gauge transformations gives a finite-dimensional  moduli space of solutions.  One can get many more interesting hyperk\"ahler manifolds from these.   Imposing an invariance condition in one direction gives the moduli space of monopole solutions to the Bogomolny equations, in two directions the moduli space of Higgs fields and solutions to the Hitchin equations, in three directions complex co-adjoint orbits and solutions to the Nahm equations.

\subsection{Hodge theory and  $\mathbf P_{tw}^1$}

In Hodge theory, one studies extra structure on the cohomology groups $H^n(M,\mathbf C)$ arising from looking at harmonic forms representing de  Rham cohomology classes.   When $M$ is a smooth compact K\"ahler manifold,  harmonic $n$-forms decompose into harmonic $(p,q)$-forms and one has a Hodge decomposition
$$H^n(M,\mathbf C)=\oplus_{p+q=n}H^{p,q}(M)$$
This decomposition does not vary holomorphically with changes in the complex structure of $M$, but the filtration on $V=H^n(M,\mathbf C)$ using the degrees of holomorphic differential forms does have this property.   

The added structure that occurs in Hodge theory is described by defining a pure real Hodge structure of weight $w$ on a complex vector space $V$ with a real structure to be a filtration
$${0}\subset F^n\subset F^{n-1}\subset \cdots \subset F^1\subset F^2=V$$
such that the filtration and its conjugate satisfy
$$F^p\oplus \overline{F^{n-p+1}}=V$$
There are more general \lq\lq mixed" Hodge structures, which we will not consider here.
Taking $V=H^n(M,\mathbf C)$ with its usual real structure
and $F^p=\oplus_{i>p} H^q(M,\Omega^i)$, one has $H^{p,q}(M)=F^p\cap\overline{F^q}$.

In \cite{simpson-mixed-twistor} and elsewhere, Simpson has shown that one can replace this definition of a Hodge structure by a definition of a \lq\lq twistor structure", which uses bundles on $\mathbf P_{tw}^1$.   The classification of holomorphic vector bundles on $\mathbf {CP}^1$ is well-known: they are direct sums of the $\mathcal O(w)$ which are degree $w$ rank $1$ line bundles.  The corresponding objects in the case of $\mathbf P_{tw}^1$ (which we'll continue to refer to as holomorphic bundles) are rank two bundles when the degree is odd, rank one when the degree is even.  They can be denoted 
$$\mathcal O_{\mathbf P^1_{tw}}\left(\frac{w}{2}\right)$$
for $w$ an integer.  Here $w/2$ is the slope (degree/rank) of the vector bundle.

Simpson identifies a real Hodge structure, pure of weight $w$, with the twistor stucture given by the holomorphic vector bundle $\mathcal E$ on $\mathbf P_{tw}^1$ such that 
$$\mathcal E=\mathcal O_{\mathbf P_{tw}^1}\left(\frac{w}{2}\right)\oplus \mathcal O_{\mathbf P_{tw}^1}\left(\frac{w}{2}\right)\oplus\cdots\oplus \mathcal O_{\mathbf P_{tw}^1}\left(\frac{w}{2}\right)$$
where the number of terms in the sum is the dimension of the underlying vector space $V$.  The filtration part of the Hodge structure corresponds to an action of the group $U(1)$ on the bundle $\mathcal E$.

To go back and forth between Hodge structures and twistor stuctures, one uses the coordinate $z$ on $\mathbf P^1_{tw}$, which picks out a point $\infty\in\mathbf P^1_{tw}$, about which one can use the coordinate $\lambda=1/z$.  A $U(1)$ group now acts on the $\mathbf P^1_{tw}$, preserving $\infty$.  Given a Hodge structure on $V$, one constructs a twistor structure by taking the trivial vector bundle 
$$\mathcal E_1=V\otimes_{\mathbf R}\mathcal O_{\mathbf P_{tw}^1}$$
away from $\lambda=0$ and gluing it to a bundle $\mathcal E_2$ on the formal neighborhood of $\lambda=0$ using the filtration. 

Given a $U(1)$ equivariant twistor structure $\mathcal E$, one can recover the Hodge structure by taking $V$ to be the fiber of $E$ at a point away from $\lambda=0$.  The action of $U(1)$ at $\lambda=0$ will decompose the fiber there into $U(1)$-weight spaces which will be the associated graded spaces of the filtration.  One can make the following table:
\begin{center}
\begin{tabular}{ |p{2.2in}|p{2.2in}| }
\hline
Hodge structures & twistor structures \\
\hline
Real Hodge structure& $U(1)$ equivariant holomorphic vector bundle $\mathcal E$ on $\mathbf P_{tw}^1$\\
\hline
pure of weight $w$& sum of $\mathcal O_{\mathbf P^1_{tw}}\left(\frac{w}{2}\right)$\\
\hline
underlying vector space $V$& trivial bundle $\mathcal E_1=V\otimes \mathcal O_{\mathbf P_{tw}^1}$ away from $\infty$\\
\hline
Hodge filtration& modification of $\mathcal E_1$ at $\infty$ to give $\mathcal E$.\\
\hline
\end{tabular}
\end{center}

\subsection{Non-abelian Hodge theory}

There is a non-abelian version of Hodge theory due to Simpson (see for instance \cite{simpson-icm}), which is based on considering for a space $X$ the moduli space of representations of the fundamental group $\pi_1(X)$.  One takes the space of representations of $\pi(X)$ into $GL(n,\mathbf C)$ up to equivalence, and denotes it
$$M=H^1(X,GL(n,\mathbf C))$$
thinking of it as non-abelian cohomology in degree one.  This is a set, not a vector space, and the non-abelian nature of the coefficients implies it does not extend to higher-degree cohomology.

For the case $X=\Sigma$ a Riemann surface, this moduli space is hyperk\"ahler, and one can study it by using the corresponding twistor space.   For different points on $\mathbf{CP}^1$ one gets different definitions of the space $M$ of quite different natures, relating moduli spaces of Higgs bundles on $\Sigma$ and moduli spaces of holomorphic flat connections on $\Sigma$ to moduli spaces of representations.  One method for constructing this twistor space uses Deligne's $\lambda$-connections, with $\lambda$ a complex parameter corresponding to a coordinate on $\mathbf P^1$.  For $\lambda=1$ one gets usual connections, at $\lambda=0$ Higgs bundles.

\section{The twistor $\mathbf P^1$ and arithmetic}

The story of the twistor $\mathbf P^1$ has an arithmetic generalization, with a different but very similar structure occuring for each prime number $p$.   For each such $p$ one can define a norm  on the field  $\mathbf Q$ of rational numbers by taking
$$|x|_p=p^{-a}$$
where $a$ is the power of $p$ that occurs when one factors $x=n/m \in \mathbf Q$.  In this norm two rational numbers are close together when their difference has a large positive factor of $p$.   Just as one defines the field $\mathbf R$ as the completion of $\mathbf Q$ with respect to the usual norm $|\cdot |$, one can define new fields $\mathbf Q_p$ as the completions of $\mathbf Q$ with respect to the norms $|\cdot |_p$.  The usual norm the corresponds to an \lq\lq infinite prime" and is denoted by $|\cdot |_\infty$.

One can then study elements of $\mathbf Q$ as elements of the completions $\mathbf Q_p$ and $\mathbf R$, and field extensions of $\mathbf Q$ in terms of field extensions of $\mathbf Q_p$ and of $\mathbf R$.   Field extensions of $\mathbf R$ are very simple to understand, since the Galois group $Gal(\mathbf C/\mathbf R)$ of the algebraic closure is just $\mathbf Z/2\mathbf Z$.  Field extensions of $\mathbf Q_p$ are a much more complicated subject, with $Gal(\overline{\mathbf Q}_p/\mathbf Q_p)$ having a rather intricate structure.

\subsection{Analogs at $p$ of $\mathbf H$ and the twistor $\mathbf P^1$}

For each prime $p$, part of the structure of $\mathbf  Q_p$ and its extensions can be understood as closely analogous to what happens at the infinite prime, in terms of generalizations of the algebra $\mathbf H$ over $\mathbf R$ to more general quaternion algebras for each $\mathbf Q_p$.  To get this generalization, for an arbitrary field $F$, consider the algebra over $F$ with basis elements $1, \mathbf i,\mathbf j,\mathbf k$ satisfying the usual quaternion relations
$$\mathbf i\mathbf j=-\mathbf j\mathbf i =\mathbf k$$
as well as the more general
$$\mathbf i^2=a,\ \ \mathbf j^2=b$$
where $a,b$ are invertible elements in $F$.  We'll use the notation
$$\left({{a,b}\over F}\right)$$
to denote this algebra.  

In this notation the usual quaternion algebra is
$$\mathbf H=\left({{-1,-1}\over \mathbf R}\right)$$
For different values of $a,b$, $\left({{a,b}\over \mathbf R}\right)$ will be isomorphic to either $\mathbf H$ or the algebra $\mathbf M(2,\mathbf R)$ of two-by-two real matrices.  These two algebras are not isomorphic, but become so after allowing complex coefficients, i.e.
$$\mathbf H\otimes_\mathbf R \mathbf C=M(2,\mathbf R)\otimes_\mathbf R \mathbf C=M(2,\mathbf C)$$

Something very similar happens for $F=\mathbf Q_p$:  the $\left({{a,b}\over \mathbf Q_p}\right)$ fall into two isomophism classes, one of which includes
$$M(2,\mathbf Q_p)$$
and the other of which includes
$$\left ({{p,u}\over \mathbf Q_p}\right)$$
where $u$ is a unit and not a square.  The second of these is  a division algebra and is the analog at the prime $p$ of the usual quaternion algebra $\mathbf H$ at the infinite prime.    Allowing coefficients in a quadratic extension $K$ of $\mathbf Q_p$ (for $p\neq 2$ there are three of these: $\mathbf Q_p[\sqrt{p}],\mathbf Q_p[\sqrt{u}],\mathbf Q_p[\sqrt{up}]$) one finds that the two non-isomorphic algebras become isomorphic
$$\left({{p,u}\over \mathbf Q_p}\right)\otimes_{\mathbf Q_p}K=M(2,\mathbf Q_p)\otimes_{\mathbf Q_p}K=M(2,K)$$
The Hilbert symbol distinguishes the two isomorphism classes, and is defined by
$$(a,b)_F=
\begin{cases} 1\ \ \text{if}\ \ \left({{a,b}\over F}\right)\simeq M(2,F)\\
-1 \ \ \text{if}\ \ \left({{a,b}\over F}\right)\not\simeq M(2,F)
\end{cases}$$

One can associate to each quaternion algebra the equation of a conic
$$\left({{a,b}\over F}\right)\leftrightarrow C=\{(x,y,z):-ax^2-by^2+abz^2=0\}$$
This will have solutions in $F$ when $(a,b)_F=1$, but not when $(a,b)_F=-1$
For the case of the quaternions $\mathbf H$ the equation for $C$ is
$$x^2+y^2+z^2=0$$
which has no non-zero solutions over the real numbers, but over the complex numbers describes $\mathbf {CP}^1\subset \mathbf {CP}^2$ in homogeneous coordinates.
At a prime $p$, the analog of the twistor $\mathbf P^1$ will be the conic $C$ with equation
$$-px^2-uy^2 +upz^2=0$$
which has no solutions in $\mathbf Q_p$, but does have solutions in the quadratic extensions $K$ of $\mathbf Q_p$.  The Galois group 
$$Gal(K/\mathbf Q_p)=\mathbf Z/2\mathbf Z$$
acts on these solutions with no fixed points, providing an analog of the antipodal map on the $\mathbf {CP}^1$.

A generalization of the class of quaternion algebras over a field $F$ is the class of finite dimensional simple algebras over $F$ with center $F$.  One can put an equivalence relation 
$$A\sim B \leftrightarrow A\otimes M_n(F)\simeq B\otimes M_m(F)$$
on such algebras, and a product 
$$[A]\cdot[B]=[A\otimes B]$$
on the equivalence classes, giving a group called the Brauer group $Br(F)$.  For $F=\mathbf R$ one can show that $Br(\mathbf R)=\mathbf Z/2\mathbf Z$, with generator $[\mathbf H]$.  

For the case $F=\mathbf Q_p$, the class $[\left ({{p,u}\over \mathbf Q_p}\right)]$ is of order two in $Br(\mathbf Q_p)$, but in this case the Brauer group is
$$Br(\mathbf Q_p)=\mathbf Q/\mathbf Z$$
with generators of order $n$ subgroups given by higher dimensional algebras.    One has higher dimensional analogs of the twistor $\mathbf P^1$, given by Brauer-Severi varieties, which are varieties over $\mathbf Q_p$ which become isomorphic to projective space when one extends scalars to the algebraic closure of $\mathbf Q_p$.

One way of understanding why one gets the same classification for quaternion algebras and these conics is that they both represent elements in the same Galois cohomology group
$$H^1(Gal(\overline F/F),PGL_2(\overline F))$$
which has two elements.  In the case $F=\mathbf R$ one gets nothing new by going to higher degree and considering
$$H^1(Gal(\overline F/F),PGL_n(\overline F))$$
but for $F=\mathbf Q_p$ one gets something new at each $n$.

In the classification of quadratic forms over a field, the Brauer group of the field (or at least its 2-torsion) makes an appearance as an invariant of quadratic forms.  The Witt ring of equivalence classes of quadratic forms up to hyperbolic planes has a $\mathbf Z/2\mathbf Z$-valued rank map, with kernel $I$.  The two-torsion of the Brauer group is isomorphic to $I^2/I^3$.

Generalizing from the case of $Br(\mathbf R)=\mathbf Z/2\mathbf Z$ in another direction, one can consider $\mathbf Z/2\mathbf Z$-graded algebras and  define in the same way as the Brauer group  instead the Brauer-Wall or super-Brauer group $SBr(F)$ of a field.   Clifford algebras provide representatives of elements of the super-Brauer group.  In particular, for $F=\mathbf R$ one finds that $SBr(\mathbf R)=\mathbf Z/8\mathbf Z$, with representatives given by the real Clifford algebras $\text{Cliff}(r,s,\mathbf R)$ for real vector spaces with a non-degenerate quadratic form of signature $(r,s)$.

The structure of $SBr(\mathbf R)$ explains the $8$-fold periodicity found in the structure of real Clifford algebras (and in the $KO$ version of K-theory).   The real Clifford algebras are given by a somewhat complicated pattern of different matrix algebras over $\mathbf R,\mathbf C,\mathbf H$, which become isomorphic after complexification:
$$\text{Cliff}(r,s, \mathbf R)\otimes_\mathbf R\mathbf C=
\begin{cases}
M(2^n,\mathbf C) \ \ \text{for}\ r+s=2n\\
M(2^n,\mathbf C)\oplus M(2^n,\mathbf C) \ \ \text{for}\ r+s=2n+1
\end{cases}$$
For details of this story, see \cite{deligne}.

\subsection{Canonical quantization, quaternions and arithmetic}

For one degree of freedom, canonical quantization starts with a position space $\mathbf R$ with coordinate $q$ and a dual momentum space $\mathbf R^*$ with coordinate $p$.  It gives operators $Q,P$ on a space of functions of $q$, which we can take to be the Schwarz space $\mathcal S(\mathbf R)$.  These operators satisfy the Heisenberg commutation relations
$$[Q,P]=i\mathbf 1$$
(we are setting $\hbar=1$) so provide an irreducible unitary representation of the Heisenberg Lie algebra $\mathfrak h_3=\mathbf R \oplus \mathbf R^*\oplus \mathbf R$ on the state space $\mathcal H=\mathcal S(\mathbf R)$.  This exponentiates to a representation of the Heisenberg Lie group $H_3$, which is unique up to unitary equivalence (by the Stone-von Neumann theorem).  The symplectic group $Sp(2,\mathbf R)=SL(2,\mathbf R)$ acts by automorphisms on the Heisenberg group, and this implies that the irreducible representation of $H_3$ provides a projective unitary representation of $SL(2,\mathbf R)$.  This is known by various names, for concision we'll refer to it as the Weil representation.  For a detailed explanation of all this,  including the generalization to any finite number of degrees of freedom, see \cite{woit-qm}.

One can replace the field $F=\mathbf R$ used here by more general local fields, for instance $F=\mathbf Q_p$.   The Weil representation in all these cases becomes a true (not just projective) representation of a non-trivial double cover of the symplectic group, called the metaplectic group.   In \cite{weil}, Weil studied these representations and the $\pm 1$-valued cocycle which determines the double cover.  He also noted that, given a vector space $V$ over $F$ of dimension $d$ with a quadratic form $Q$, one could replace the symplectic vector space $F\oplus F^*$ by the symplectic vector space $W=(F\oplus F^*)\otimes V$ and again construct a Weil representation, which now will be an irreducible projective representation of $Sp(2d,F)$.  If one takes $V$ to be a quaternion algebra $\left({{a,b}\over F}\right)$
and writes
$$W_{a,b}=(F\oplus F^*)\otimes\left({{a,b}\over F}\right)$$
 then Weil showed that the Brauer class
$$\left[\left({{a,b}\over F}\right)\right]\in \mathbf Z/2\mathbf Z$$
determines whether the Weil representation constructed using $W_{a,b}$ is a true or projective (up to a factor $\pm 1$) representation of the symplectic group.

The construction of the metaplectic extension of the symplectic group is a rather intricate story, even for the local field $\mathbf R$, see for example \cite{lion-vergne}.  The extension can be constructed using the Maslov index, which takes values in the Witt group of a quadratic form.   The metaplectic extension just depends on $I^2/I^3$, the piece of this given by two-torsion in the Brauer group.

It is rather striking that one gets in this way a generalization to the arithmetic context of a theory which for $F=\mathbf R$ is canonical quantization for the case of configuration space $\mathbf H$, which is just the case of quantization relevant to wave-functions on the usual four-dimensional space-time.  The author is not aware of this being previously considered from any physical viewpoint, but it clearly deserves further study.

So far we have just been considering the local fields $\mathbf Q_p$ and $\mathbf R$, but one can use these to study the global field $\mathbf Q$.  If one takes $a,b$ invertible elements in $\mathbf Q$, one finds that the Hilbert symbols of the corresponding $a,b$ for the various primes satisfy
$$\prod_{p}(a,b)_p=1$$
where the product is over the finite primes (for which $(a,b)_p=(a,b)_{\mathbf Q_p}$) and the infinite prime (for which $(a,b)_{\infty}=(a,b)_\mathbf R$). This is known as the Hilbert reciprocity law, and from it one can easily derive Gauss's quadratic reciprocity law.    

In \cite{weil} Weil shows that one can construct Heisenberg and metaplectic groups, as well as the Weil representation, not just for the local fields $F=\mathbf Q_p,\mathbf R$, but also for the ad\`ele group $\mathbf A_\mathbf Q$ which puts all the local fields together.   If one does this for $W_{a,b}$,
replacing $F$ by $\mathbf A_{\mathbf Q}$,  one can show that this ad\`elic Weil representation is a true (rather than projective) representation of the diagonal $Sp(W_{a,b},\mathbf Q)$ subgroup of the corresponding adelic symplectic group and that this fact is equivalent to Hilbert reciprocity.

For more about this, besides \cite{weil}, see \cite{howe} and \cite{gelbart}.  Howe showed that in general one can get a lot of mileage out of  forming the symplectic vector space $W=U\otimes V$ for $U$ a symplectic vector space and $V$ an orthogonal vector space, taking the Weil representation for $W$, then restricting this to the subgroup $Sp(U)\times O(V)$.  One gets in this way (\lq\lq Howe duality") a matching between irreducible representations of $Sp(U)$ and of $O(V)$ and in the ad\`elic context  a \lq\lq theta correspondence" between automorphic forms.  Jacquet and Langlands used the case considered above of $U=F\oplus F^*$ and $V$ a quaternion algebra in their demonstration of an example of Langlands functoriality.

Witten in \cite{witten-weyl} and \cite{witten-automorphic} (see also \cite{beilinson-feigin-mazur}) discussed a formulation of holomorphic conformal field theories such as the theory of a free fermion on a Riemann surface in terms of a function field (rather than number field) analog of the above.  Instead of the Hilbert symbol, one has an analogous symbol due to Weil, and there's an analogous reciprocity law.

\subsection{The Fargues-Fontaine curve and the twistor $\mathbf P^1$}

The twistor $\mathbf P^1$ turns out to be the analog at the infinite prime of a more sophisticated arithmetic geometry version of a curve (called the \lq\lq Fargues-Fontaine curve") which can be defined for each prime $p$.  The Fargues-Fontaine curve provides a geometrical interpretation of some of the structure of $p$-adic Hodge theory.  It has recently been used by Fargues and Scholze\cite{fargues-scholze} to recast the arithmetic local Langlands conjecture in terms of a geometric Langlands conjecture on the curve.

{\bf Warning}:{\it this section is about an analogy between two different things, but I only actually understand one of them.  To limit my embarassment I'm just giving a comparison table.  The reader should pay minimal attention to what is here, and instead consult the following sources explaining this material, written by people who understand both sides of the analogy.}

\begin{itemize}
\item
Overview of the Fargues-Fontaine curve by Jacob Lurie:

\url{https://www.math.ias.edu/~lurie/ffcurve/Lecture1-Overview.pdf}
\item
Lectures by Laurent Fargues at Salt Lake City

\url{https://webusers.imj-prg.fr/~laurent.fargues/SaltLake.pdf}

and Beijing

\url{https://webusers.imj-prg.fr/~laurent.fargues/Course%20Shenxing.pdf}

as well as seminars at Columbia

\url{https://webusers.imj-prg.fr/~laurent.fargues/padic_Twistors.pdf}

and Jussieu

\url{https://webusers.imj-prg.fr/~laurent.fargues/Twisteurs_p_adiques.pdf}

\item
Peter Scholze's 2018 ICM talk on p-adic geometry \cite{scholze-icm}

\url{https://arxiv.org/abs/1712.03708}

See in particular section 6 and the top of Figure 3.

\end{itemize}

\begin{center}
\begin{tabular}{ |p{2.2in}|p{2.2in}| }
\hline
finite $p$ & infinite $p$\\
\hline
\hline
$\mathbf Q_p$ & $\mathbf R$\\
\hline
$\mathbf C_p$, completion of $\overline{\mathbf Q}_p$ & $\mathbf C$\\
\hline
Fargues-Fontaine curve $FF_p$ & $\mathbf P_{tw}^1$\\
\hline
pt. at $\infty$ given by $i:Spec\ \mathbf C_p\to FF_p$&  $0,\infty\in \mathbf {CP}^1$, $\infty\in \mathbf P_{tw}^1$\\
\hline
$Gal(\overline{\mathbf Q}_p/\mathbf Q_p)$ action on $FF_p$& $U(1)$ action on $\mathbf P_{tw}^1$\\
\hline
vector bundles on $FF_p$& vector bundles on $\mathbf P_{tw}^1$\\
\hline
vector bundles classified by fractions & vector bundles classified by half-integers\\
\hline
$Gal(\overline{\mathbf Q}_p/\mathbf Q_p)$-equivariant vector bundles on $FF_p$& $U(1)$-equivariant vector bundles on $\mathbf P_{tw}^1$\\
\hline
$p$-adic Hodge structure & Hodge structure \\
\hline
Fontaine ring $B^+_{dR}=\widehat{\mathcal O}_{FF_p,\infty}$& Power series ring $\mathbf C[[\lambda]]=\widehat{\mathcal O}_{\mathbf P^1_{tw},\infty}$\\
\hline
$\widehat{\mathbf Z}$-cover of $FF_p$&$ \mathbf{CP}^1$ is $\mathbf Z/{2\mathbf Z}$-cover of $\mathbf P^1_{tw}$\\
\hline
\end{tabular}
\end{center}

\section{Speculations}

The significance of the occurrence of the same twistor $\mathbf P^1$ as a fundamental description of a point in space-time physics and in number theory remains obscure.    David Ben-Zvi has emphasized (see for instance here \cite{ben-zvi-langlands}) that one can take a four-dimensional topological quantum field theory point of view on number theory, in the context of well-known analogies relating number fields to 3-manifolds and primes to knots \cite{knotsandprimes}.    Kapustin-Witten \cite{kapustin-witten} realize the geometric Langlands program duality in terms of electro-magnetic duality in such a $4$d quantum field theory.    In this quantum field theory, the self-duality equations play a central role, and the Penrose-Ward transform indicates that twistor theory is the geometric framework for understanding the significance of these equations.  

It may very well be that the appearance of the  twistor $\mathbf P^1$ in physics and in number theory is of no fundamental significance, but it also seems quite possible that it is pointing towards some insight relating the two subjects yet to be found.

\section{Acknowledgements}

Many thanks to Johan de Jong for his help with the material of these notes,  to Peter Scholze for pointing out to me that the same twistor $\mathbf P^1$ of Euclidean twistor theory was showing up as an analog of the Fargues-Fontaine curve at the infinite prime, and to David Ben-Zvi and Brian Conrad for helpful comments.

\printbibliography[title={References}]
\end{document}